%% file: paper.tex
\documentclass[letterpaper,twocolumn,10pt]{article}
\usepackage{usenix-2020-09}
\let\ACMmaketitle=\maketitle
\renewcommand{\maketitle}{\begingroup\let\footnote=\thanks \ACMmaketitle\endgroup}
\usepackage{tikz}
\usepackage{amsmath}
\usepackage{flushend}

\begin{document}

\date{}

\def\Section {\S}

\sloppy


\title{\Large \bf Sustainable Computing -- Without the Hot Air\footnote{Title inspired by the book Sustainable Energy - Without the Hot Air~\cite{mackay2008sustainable}.}}

\author{
{\rm Noman Bashir, David Irwin, Prashant Shenoy, Abel Souza}\\
University of Massachusetts Amherst
} 

\maketitle

\begin{abstract}
\input{abstract.tex}

\end{abstract}

\section{Introduction}
\input{introduction.tex}
\label{sec:introduction}

\input{efficiency}
\bibliographystyle{plain}
\bibliography{paper}

\end{document}

%% file: abstract.tex

The demand for computing is continuing to grow exponentially. This growth will translate to exponential growth in computing's energy consumption unless improvements in its energy-efficiency can outpace increases in its demand.  Yet, after decades of research, further improving energy-efficiency is becoming increasingly challenging, as it is already highly optimized.  As a result, at some point, increases in computing demand are likely to outpace increases in its energy-efficiency, potentially by a wide margin. Such exponential growth, if left unchecked, will position computing as a substantial contributor to global carbon emissions.  While prominent technology companies have recognized the problem and sought to reduce their carbon emissions, they understandably focus on their successes, which has the potential to inadvertently convey the false impression that this is now, or will soon be, a solved problem.  Such false impressions can be counterproductive if they serve to discourage further research in this area, since, as we discuss, eliminating computing's, and more generally society's, carbon emissions is far from a solved problem. To better understand the problem's scope, this paper distills the fundamental trends that determine computing's carbon footprint and their implications for achieving sustainable computing.

%% file: introduction.tex
\vspace{-0.3cm}
The demand for computing is growing exponentially, and has been for some time~\cite{cacm-article}, mostly because society continues to find useful applications for it.  These applications have transformed modern society over the past fifty years, and now largely serve as the foundation of its information-based economy.  Since computation is simply a processed form of energy, and energy use incurs both a monetary and environmental cost, there has long been a concern that exponentially growing computing demand would translate into exponentially growing energy demand, which could stifle technological innovation and damage the environment.  For example, as early as 2007, the U.S. Environmental Protection Agency (EPA) sent a report to Congress projecting a doubling in aggregate data center energy consumption if historical trends continued~\cite{us2007epa}.  While follow-up analyses in 2011~\cite{koomey11} and 2016~\cite{lbnl} suggested the 2007 forecast was inaccurate (with data center energy consumption increasing by only 24\% from 2005-2010 and 4\% from 2010-2014), more recent studies have come to widely different estimates of the growth in data center energy consumption~\cite{hintemann2020efficiency, andrae2019projecting, andrae2015global, belkhir2018assessing, shift-project}.  The most optimistic of these analyses estimates only a 6\% increase in data center energy consumption from 2010-2018, or roughly an average 0.65\% increase per year, despite a 6$\times$ increase in capacity~\cite{masanet}. 

Ultimately, computing's aggregate energy consumption $E$ (in kWh) is simply a function (shown below) of its demand $d$ (in cycles) versus its energy-efficiency $e$ (in cycles/kWh). 

\vspace{-0.3cm}
\begin{equation}
E = \frac{d}{e}
\label{equation:energy}
\end{equation}

Likewise, the growth rate $r_E$ in computing's energy consumption is simply a function of the growth rate in its demand ($r_d$) versus its energy-efficiency ($r_e$): if demand increases faster than energy-efficiency, i.e., $r_d > r_e$, then consumption will grow exponentially, and otherwise, it will shrink. 

The optimistic analysis above attributes the small increase in computing's energy consumption to the incredible increase in computing's energy-efficiency over the past 15 years.  This trend largely derives from computing's ongoing transition from smaller traditional data centers to ``hyperscale'' cloud platforms, which have a strong financial incentive to optimize their energy-efficiency to reduce operational costs.  Indeed, the reported Power Usage Effectiveness (PUE) for Google's data centers---the ratio of their total energy to the energy of IT equipment---is now $\sim$$1.1$~\cite{google-pue}, which is near the optimal value of $1$ and also nearly 30\% lower than the industry average of $\sim$$1.57$~\cite{uptime}.  However, while the trends above have been broadly characterized as a tremendous success story for industry, which they undoubtedly are, they also belie a significant problem.  Specifically, \emph{despite incredible improvements in computing's energy-efficiency over the past 15 years, by even the most optimistic analysis, its aggregate energy consumption still increased!}  That is, $r_d > r_e$ over this period. 

Unfortunately, after decades of research, further improving energy-efficiency is becoming increasingly challenging, as it is already highly optimized.  Thus, moving forward, increases in computing's energy-efficiency are likely to slow, especially once its transition to the cloud is complete.  At the same time, demand is likely to continue increasing, if not accelerate, as new useful applications are developed.  For example, recent progress in AI has the potential to enable a wide range of novel applications that are also computationally-intensive~\cite{openai}.  Importantly, these trends will not only have profound implications on computing's cost, but also its carbon emissions.  Exponentially growing demand that is not offset by energy-efficiency improvements, would quickly position computing as a substantial contributor to global carbon emissions.  Yet, at the same time, there is now a broad consensus that society must rapidly reduce, and ultimately eliminate, its carbon emissions to halt climate change, which represents an existential threat to the earth's ecosystem and humanity.   

Similar to the relationship above, computing's carbon footprint $C$ (in g$\cdot$CO$_2$) is simply a function (shown below) of its aggregate energy consumption $E$ from Equation~\ref{equation:energy} versus its energy's carbon-efficiency $c$ (in kWh/g$\cdot$CO$_2$).

\vspace{-0.2cm}
\begin{equation}
C = \frac{E}{c} = \frac{d}{c \times e}
\label{equation:carbon}
\end{equation}

Likewise, the growth rate $r_C$ in computing's carbon footprint is simply a function of the growth rate in its aggregate energy consumption ($r_E$) versus its energy's carbon-efficiency ($r_c$): if consumption increases faster than carbon-efficiency, i.e., $r_E > r_c$, then it will grow exponentially, and otherwise, it will shrink.   We can also use Equation~\ref{equation:energy} to substitute $d/e$ for $E$ in Equation~\ref{equation:carbon}. Here, $c \times e$ represents computing's carbon-efficiency (in cycles/g$\cdot$CO$_2$), and highlights that computing's energy-efficiency and energy's carbon-efficiency are equally important in determining computing's carbon footprint. 


The most recent estimates suggest electricity's carbon-intensity (in g$\cdot$CO$_2$/kWh), which is the inverse of carbon-efficiency, in the U.S. decreased 30\% between 2001 and 2017, largely due to the replacement of coal-fired power plants with natural gas and wind generation~\cite{carbon-intensity,holland2022marginal}.  This is equivalent to a 45.6\% increase in energy's carbon-efficiency over the same period, or equivalently a $\sim$2.33\% increase per year. Thus, the most optimistic assessments based on the reported averages above---a 0.65\% per year increase in energy consumption~\cite{masanet} and a 2.33\% per year increase in energy's carbon-efficiency~\cite{carbon-intensity}---suggest that computing's carbon footprint decreased slightly (by $\sim$1.64\% per year on average\footnote{Carbon footprint's growth rate is (M-N)/(1+N), where M and N are the growth rates in energy's consumption and carbon-efficiency, respectively.}) from 2010-2017, and that this decrease was entirely due to improvements in energy's carbon-efficiency.   In contrast, if the most optimistic assessments are inaccurate, then, based on the same reasoning, computing's carbon footprint likely increased. 

Of course, the macro longitudinal analyses cited above are necessarily simplistic, coarse, and imprecise, and should be taken with a grain of salt, i.e., viewed skeptically.   For example, our analysis does not take into account that electricity's carbon-efficiency varies widely across days, seasons, and regions, and thus it is also a function of when and where energy is consumed.  That said, macro analyses can be useful in distilling the fundamental trends that matter.  In particular, irrespective of the specific numbers, Equation~\ref{equation:carbon} shows that the growth of computing's carbon footprint is based on the relative growth in its demand, energy-efficiency, and energy's carbon-efficiency.  Thus, better understanding these relative growth rates can provide some insight into how computing's carbon footprint is changing, and also how to eliminate it. 

Our simple analysis also paints a slightly different, and more nuanced, picture of computing's carbon footprint than recent industry announcements~\cite{amazon-carbon-neutral,facebook-carbon-neutral,google-carbon-free,microsoft-carbon-negative}. While prominent technology companies have recognized the trends above and sought to reduce their carbon emissions, they understandably focus on their successes, which has the potential to inadvertently convey the false impression that this is now, or will soon be, a solved problem. This paper's title is a reference to a well-known book that made a similar observation about the energy industry~\cite{mackay2008sustainable}. For example, many technology companies have eliminated their net carbon emissions~\cite{amazon-carbon-neutral,facebook-carbon-neutral,google-carbon-free,microsoft-carbon-negative}, which they often refer to as running on ``100\% renewable energy.''  However, eliminating net carbon emissions is both different and much easier than eliminating direct carbon emissions. Unfortunately, such false impressions can be counterproductive if they unintentionally discourage further research, since, as we discuss, eliminating computing's, and more generally society's, real carbon emissions is far from a solved problem. To better understand the problem's scope, we examine relative trends in the growth of computing's demand and energy-efficiency, as well as its energy's carbon-efficiency, and their implications for achieving sustainable computing. 

%% file: efficiency.tex


\section{Computing's Demand}

By all indications, the demand for computing---the total number of cycles executed---has been growing exponentially for some time, likely since the dawn of computing~\cite{cacm-article}.  The optimistic analysis above estimated a 6$\times$ increase in data center capacity from 2010-2018 (or $\sim$22\% per year)~\cite{masanet}.  Another recent study estimated that the capacity for the most efficient hyperscale data centers had doubled over the past five years~\cite{srgresearch20}. While some of this growth surely represents existing demand transitioning from smaller traditional data centers to cheaper cloud platforms, much of it also likely represents new demand from cloud-native applications.   For example, a recent report estimates that 75\% of companies are now focusing on developing cloud-native applications~\cite{cloud-native}.  A variety of other anecdotal evidence suggests computing demand may be accelerating. For example, the cycles devoted to cryptomining~\cite{bitcoin-energy} and training state-of-the-art machine learning (ML) models~\cite{ai-index-2019} is growing much faster than Moore's Law.  Computing is also continually displacing other activities, such as videoconferencing in lieu of traveling for meetings. While such displacement may improve energy-efficiency, which we discuss below, it undoubtedly increases computing's demand. 

The only way to reduce computing demand (aside from not computing) is to improve algorithmic efficiency by enabling computation to do more (or the same) work using fewer cycles.  To be sure, there are numerous and substantial remaining opportunities to improve algorithmic efficiency.  For example, broad adoption of proof-of-stake consensus for cryptocurrencies would effectively eliminate soaring demand from cryptomining. Likewise, reducing the demand to train large-scale ML models has been a focus of recent research, and yielded some notable improvements~\cite{ml-efficient1,ml-efficient2}. Importantly, though, computing's demand is not only a function of each applications' efficiency, but also the total number of applications executed.  That is, improving the efficiency of training ML models by 10$\times$ will not decrease demand if the number of models trained increases by 10$\times$.  Absent resource constraints, computing's potential applications still seem limitless, or at least only limited by people's imaginations.  Thus, improvements to algorithmic efficiency may be hard-pressed to offset the growth in the sheer number of applications executed.

Finally, while industry has a strong incentive to increase algorithmic efficiency to reduce their operational cost, it is bounded by each problem's computational complexity.  Obviously, we cannot solve computational problems without some minimal amount of computation.  Further, industry's primary incentive is to increase its potential profit, which is effectively unbounded and generally correlates with increasing demand, regardless of efficiency.  That is, while improving efficiency may increase profit, it is not always necessary or possible. 

\noindent {\bf Key Point.} \emph{Computing demand is increasing, and possibly accelerating, as more useful applications are developed. Improvements to algorithmic efficiency are bounded and thus unlikely to staunch this growth over the long-term.}

\section{Computing's Energy-Efficiency}

Computing's energy-efficiency has also been increasing at an exponential rate for some time, a trend that has been referred to as Koomey's Law~\cite{koomey1}.   Koomey estimated that computing's energy-efficiency at peak capacity has been approximately doubling every 1.57 years from the 1950s up through 2010 (roughly in-line with Moore's law)~\cite{koomey1}, although a revised analysis suggested the pace slowed to every 2.6 years starting in 2000 (due, in part, to the end of Dennard scaling)~\cite{revised-koomey}.  Since computing platforms are often idle, the same report also estimated that average energy-efficiency, which considers idle periods, had continued to double every $\sim$1.5 years due to increases in average utilization and energy-proportionality.   This latter point captures some of the energy-efficiency improvements from the transition to cloud platforms, which leverage statistical multiplexing at massive scales to increase average server utilization, as servers are more energy-efficient at higher utilization.  As noted earlier, hyperscale cloud data centers have also improved their facilities' energy-efficiency by driving down their PUEs to within 10\% of optimal~\cite{google-pue}.

The optimistic analysis from \Section\ref{sec:introduction} estimated that the energy-efficiency improvements above have nearly kept computing's energy consumption constant, despite its exploding demand~\cite{masanet}.  Indeed, the transition from highly inefficient small traditional data centers to highly efficient hyperscale data centers has yielded dramatic increases in energy-efficiency.  Moreover, this transition is not yet complete with recent estimates suggesting nearly 20\% of data center energy consumption still derives from traditional smaller, and less efficient, facilities, so there is still room for further improvement~\cite{iea-growth-by-type}.  

Yet, continuing to increase computing's energy-efficiency to keep pace with increases in demand may prove challenging for many reasons.   Most importantly, the shift to hyperscale cloud data centers, which has yielded much of the improvement above, is a one-time event.  Once the shift is complete, it is unclear where significant improvements will come from.  One possibility is increasing the use of specialized hardware, which is more energy-efficient than general-purpose platforms. For example, cryptomining and ML have employed hardware specifically tailored to their function to dramatically increase their energy-efficiency (and performance).  However, much of computing's demand remains general-purpose, with specialized tasks still constituting only a small fraction of it.  For example, a recent paper estimates that only 15\% of Google's energy consumption is due to ML~\cite{arxiv-google}. More generally, improving computing's energy-efficiency has been a significant focus of research for at least three decades. Thus, there are likely few remaining substantial optimization opportunities using traditional methods, which may be one reason for the reported slowing of Koomey's law~\cite{revised-koomey}.

As with algorithmic efficiency, there is also a well-known physical limit to the energy-efficiency of our current form of computing, which is defined by Landaur's principle~\cite{reversible}.   Current estimates are that if computing's energy-efficiency were to continue to double every $\sim$1.5 years, then it would reach this physical limit by 2050~\cite{koomey1}, although it is not yet known how close CMOS circuits can, in practice, come to this limit.  While, in theory, adopting reversible computing techniques can overcome Landaur's limit by performing computation without consuming any energy,  it is a nascent, and largely theoretical, area that is far from any practical application~\cite{reversible1}.

Finally, even if computing's energy-efficiency were to continue doubling every 1.5 years, there is no guarantee it would cause computing's energy consumption to decrease.  As noted earlier, even by the most optimistic estimates, computing's incredible energy-efficiency improvements have not reduced its energy consumption thus far.  Interestingly, whether increases in energy-efficiency actually decrease energy's consumption is, in part, a function of economics. Specifically, as computing's energy-efficiency improves, its energy cost generally decreases, which in-turn affects its demand.  The magnitude of this effect is a function of computing's price elasticity of demand, which dictates how much demand changes when prices change.  Jevons Paradox, which is well-known in energy economics, occurs when demand elasticity is high enough that the increases in energy consumption from higher demand (caused by lower costs) is greater than the decrease in consumption from improved energy-efficiency~\cite{jevons1,jevons2}.  Thus, under Jevon's Paradox, improved energy-efficiency actually, and paradoxically, can lead to increased energy consumption.  Even if Jevons Paradox does not occur, assessing the effect of increases in computing's energy-efficiency on its energy consumption is largely an economic, and not technical, question. 

Of course, improving computing's energy-efficiency is always beneficial, as it increases productivity and economic output, i.e., enables more to be done with less energy at lower cost.  Thus, as with improving algorithmic efficiency, industry has a strong financial incentive to improve energy-efficiency. This incentive has likely driven the incredible energy-efficiency improvements over the past fifty years.  However, improvements in computing's energy-efficiency do not necessarily, and have not historically, led to reductions in its energy consumption. In fact, if Jevons Paradox occurs, improving computing's energy-efficiency may contribute to increasing energy consumption.  

\noindent {\bf Key Point.} \emph{Computing's energy-efficiency is continuing to increase, although its rate may be slowing.   Improvements to computing's energy-efficiency are bounded, and do not necessarily, and have not historically, led to reductions in computing's energy consumption, due to faster growth in demand both from new applications and lower costs.}

\section{Energy's Carbon-Efficiency}

Unlike algorithmic- and energy-efficiency, there is no fundamental limit to energy's carbon-efficiency, since it is possible to use zero-carbon energy sources, such as solar, wind, geothermal, hydroelectric, nuclear, etc.  The cost for solar and wind renewable energy sources, in particular, have also been decreasing exponentially for some time.  Swanson's law, which captures this trend for solar energy, refers to the observation that solar photovoltaic PV module prices have tended to drop 20\% for every doubling in production volume~\cite{swanson}.  As a result, solar energy's cost (in \$/watt) has dropped $\sim$10\% each year on average over the past fifty years~\cite{solar-cost}.  Renewable energy sources also have massive energy potential that could fuel exponential growth for the foreseeable future.  For example, the amount of solar energy the earth receives each hour is more than global annual energy consumption~\cite{solar-stat}. 

As mentioned in \Section\ref{sec:introduction}, energy's carbon-efficiency has been steadily increasing for the past 20 years, mostly due to the adoption of natural gas and wind.  This trend has been independent of any efforts by the computing industry to reduce its carbon footprint.  However, isolating and capturing the trend the carbon-efficiency of computing's energy is more challenging, as it depends on the strictness of carbon accounting and attribution methods used.  Carbon offsets are the loosest, and most widely used, method of carbon accounting: they enable ``offsetting'' the use of carbon-intensive grid energy with zero-carbon renewable energy generated at another location and time.  Technology companies have led in the adoption of carbon offsets, and many have used them to eliminate their net carbon footprint, which is often referred to as running on 100\% renewable energy~\cite{amazon-carbon-neutral,facebook-carbon-neutral,google-carbon-free}.   However, while carbon offsets are beneficial in subsidizing renewable energy, they are only a temporary mechanism as society transitions to lower carbon energy, since near zero-carbon there will not be any carbon left to offset.  In addition, the use of carbon offsets means that even net zero companies are still responsible for a significant amount of direct carbon emissions. 


To reach zero-carbon, companies must progressively adopt stricter forms of carbon accounting.  To this end, Google recently announced that it aims to be ``carbon free'' by 2030,  in part, by piloting a stricter form of carbon offset, called Time-based Energy Attribute Certificates (TEACs), which have an hourly location-specific accounting regime~\cite{google-cfe}.  However, TEACs are still carbon offsets, just at a higher temporal and spatial resolution than typical offsets, which are usually 1 year and the entire earth, respectively.  That is, TEACs match consumption of grid energy within an hour to renewable generated that hour within the same grid. Thus, while TEACs are an improvement upon existing annualized location-agnostic carbon offsets, they, by definition, also cannot be used to reach zero-carbon.  Of course, since the grid cannot physically isolate different energy sources, in reality, all loads that consume grid energy share in its carbon emissions.  Thus, the strictest form of carbon accounting attributes the grid's carbon emissions to all its loads based on their energy use.   As a result, reducing and ultimately eliminating computing's carbon emissions will require changing its operations to be responsive to variations in grid energy's carbon emissions and availability. 

Thus far, we have focused on trends in operational carbon, i.e., carbon emissions from using grid energy.  There has also been an increasing focus on accounting for and reducing ``embodied carbon,'' which represents the carbon emissions from producing a product or service~\cite{chasing,embodied2,embodied3}.  For example, computing's embodied carbon emissions are based on the carbon emissions from manufacturing the facilities and IT equipment that host it.   Importantly, though, one company's embodied carbon is another company's operational carbon.  For example, a cloud platform's embodied carbon is, in part, a chip manufacturer's operational carbon.  The primary purpose in accounting for embodied carbon is to provide an incentive in the supply chain for companies to reduce their operational carbon.  That is, if companies made purchasing decisions to reduce their embodied carbon, it would incentive upstream suppliers to in-turn reduce their operational carbon.  Accounting for embodied carbon is akin to a value added tax (VAT), as carbon emissions, similar to a VAT, are associated with the value added at each production stage of a good or service.   

Unfortunately, unlike with algorithmic- and energy-efficiency, there are not yet strong financial incentives for companies to reduce their operational or embodied carbon emissions, as energy prices do not yet incorporate the cost of carbon's negative externalities to the environment.   As a result, while energy's carbon-efficiency has been improving, and is unbounded, its long-term trend is unclear.  

\noindent {\bf Key Point.} \emph{Energy's carbon-efficiency is increasing, although its long-term trend is unclear due to the lack of financial incentive to improve it.  Improvements to energy's carbon-efficiency are unbounded, as it is possible to only use zero-carbon energy.  Since there will be no carbon offsets at zero-carbon, eliminating computing's carbon emissions will ultimately require eliminating the grid's carbon emissions.} 

\section{Implications for Sustainable Computing}

The trends above have important implications for sustainable computing moving forward.  Specifically, given the fundamental limits to improving computing's algorithmic- and energy-efficiency, the only way to sustain exponential growth in its demand, while also eliminating its carbon footprint, is to improve its energy's carbon-efficiency.  However, the trends in energy's carbon-efficiency are not yet clear.  In particular, the terminology above around different forms of carbon accounting, e.g., ``100\% renewable energy,'' ``carbon-neutral,'' ``carbon-free,'' ``zero-carbon,'' `embodied carbon,'' etc., is complex and fully understanding it requires some non-trivial technical background on how society's energy system works.  To anyone without such a background, which includes much of the general public as well as many computing researchers, the use of the terms above may inadvertently convey the false impression that computing's carbon emissions are already at zero, or soon will be.  Such messaging is often pejoratively referred to as ``greenwashing.''  False impressions of computing's carbon footprint are a significant issue, as they can diminish the perception of progress in decarbonizing computing, or even discourage further research altogether. 

In the end, as we discuss, the various forms of carbon accounting and offsets are temporary measures that, by definition, will not be applicable at zero-carbon.  To reach zero-carbon, computing, and more generally society, will have to significantly change how it operates to directly use renewable and low-carbon energy.  Of course, the problem with renewable energy is that, while it is potentially plentiful, cheap, and clean, it is also highly unreliable. In particular, solar and wind vary widely and uncontrollably over time based on the earth's movement and weather.   As a result, transitioning the grid to operate entirely on zero-carbon energy will require either i) significant over-provisioning within the energy system, e.g., of batteries, solar, wind, etc., which is likely cost-prohibitive, or ii) significant flexibility in the system's loads.  
 
Fortunately, compared to other loads, computing is uniquely flexible with substantial performance, temporal, and spatial flexibility, enabling it to shift the intensity, time, and location of its execution to better align with when and where renewable and other low-carbon energy is available.  To the best of our knowledge, computing is the only load with substantial spatial flexibility that is capable of migrating its energy consumption over long distances. In addition, computing can also leverage numerous software-based fault-tolerance techniques, e.g., checkpointing, replication, and recomputation, to continue execution despite unexpected renewable shortages, which may require throttling or shutting down servers.  Thus, computing has the potential to leverage its multiple dimensions of flexibility to not only lower its direct carbon footprint, but offset variations in renewable energy's availability.  As a result, computing is not just another grid load, as it can also act as an energy resource, akin to a battery, that the grid can deploy to balance demand with a variable supply~\cite{hotnets21}.  In some sense, improving energy's carbon-efficiency is related to improving computing's energy-flexibility by enabling it to adapt to when and where low-carbon energy is available.  

While many have recognized computing's unique dimensions of energy flexibility, there has been much less research on exercising them to optimize energy's carbon-efficiency compared to computing's energy-efficiency, even though, as Equation~\ref{equation:carbon} shows, carbon-efficiency is just as important as energy-efficiency in determining computing's carbon footprint.    One reason for the lack of research is likely that, unlike with algorithmic- and energy-efficiency, there is neither a direct nor strong financial incentive to improve energy's carbon-efficiency, although this may change as renewable energy prices drop.  That said, there is a weak, but increasing, indirect incentive to track and improve energy's carbon-efficiency both to appeal to environmentally-conscious consumers (and employees), and as a hedge against future changes in the energy system, such as energy constraints due to geopolitical events, stricter carbon regulations imposed by governments, or further significant drops in renewable or battery prices. 

Another reason for the lack of research may also be that optimizing carbon-efficiency requires deeper visibility into energy's carbon emissions, which has historically not been available. Recently, carbon information services, such as electricityMap~\cite{electricity-map} and WattTime~\cite{watttime}, have emerged, and are beginning to address this issue by tracking grid energy's carbon emissions for different regions over time, and making them available online.   The data shows that grid energy's carbon emissions vary significantly by region and over time.  Cloud platforms have started adopting these services to enable their users to estimate the carbon emissions of their energy consumption, and adjust their operations to reduce emissions~\cite{google-dashboard}. 

Ultimately, the primary implication for achieving sustainable computing from the trends above is that research should emphasize improvements to  the carbon-efficiency of both computing's energy (by adapting to when and where low-carbon energy is available), as well as the grid's energy (by leveraging computing as an energy resource).   The former is important for reducing computing's direct carbon emissions, while the latter is important for reducing society's carbon emissions, which are related and also affect embodied carbon.  Given the lack of a strong financial incentive to improve carbon-efficiency, academic research in this area is especially important.  Indeed, historically, an explicit purpose of academic research has been to focus on problems that industry does not address due to lack of a near-term financial incentive. 

\noindent{\bf Acknowledgements.} This research is supported by NSF grants 2105494, 2021693,  2020888, as well as VMware.